\documentclass[a4paper,12pt]{article}
\usepackage{amssymb}
\usepackage{latexsym}
\usepackage[dvips]{graphicx}
\usepackage{cite}
\newcommand{\be}{\begin{equation}}
\newcommand{\ee}{\end{equation}}
\newcommand{\bea}{\begin{eqnarray}}

\newcommand{\eea}{\end{eqnarray}}

\begin{document}

\begin{titlepage}

\begin{centering}
\vspace{.3in}
{\LARGE{\bf M{\o}ller Energy-Momentum Complex of a Static Axially
Symmetric Vacuum Space-Time}}
\\

\vspace{.5in} {\bf  Ragab M. Gad\footnote{ragab2gad@hotmail.com} }\\

\vspace{0.3in}
\end{centering}
\begin{centering}
\normalsize {Mathematics Department, Faculty of Science,}\\
\normalsize {Minia University, 61915 El-Minia,  EGYPT.}\\
\end{centering}

\vspace{0.7in}
%%%%%%%%%%%%%%%%%%%ABSTRACT%%%%%%%%%%%%%%%%%%%%%%%%%%%%%%%%%%
\begin{abstract}
The energy and momentum densities  associated with the Weyl metric
are calculated  using M{\o}ller's energy-momentum complex.
Further, using these results we obtain the energy and momentum
densities for the Curzon metric which is a particular case of the
Weyl metric. The results are compared with the results obtained by
using the energy-momentum complexes of Einstein, Landau and
Lifshitz, Papapetrou and Bergmann. We show that the aforementioned
different prescriptions and that of M{\o}ller do not give the same
energy density, while give the same momentum density.
\end{abstract}
%%%%%%%%%%%%%%%%%%%%%%%%%%%%%%%%%%%%%%%%
\end{titlepage}
\newpage

\baselineskip=18pt
%%%%%%%%%%%%%%%%%%%%%%%%%%%%%%%%%%%%%%%%%%%%%%%%%%%%%%%%%%%%%%%%%%%%%%%%%%%%%%%%%%%%%%%%%%%%%%%%%
%%%%%%%%%%%%%%%%%%%%%%%%%%%%%%%%%%%%%%%%%%%%%%%%%%%%%%%%%%%%%%%%%%%%%%%%%%%%%%%%%%%%%%%%%%%%%%%%
%%%%%%%%%%%%%%%%%%%%%%%%%%%%%%%%%%%%%%%%%%%%%%%%%%%%%%%%%%%%%%%%%%%%%%%%%%%%%%%%%%%%%%%%%%%%%%%%%
%%%%%%%%%%%%%%%%%%%%%%%%%%% INTRODUCTION %%%%%%%%%%%%%%%%%%%%%%%%%%%%%%%%%%%%%%%%%%%%%%%%%%%%%%%%
\section*{Introduction}

The problem of energy and momentum localization has been one of
the oldest but most interesting and thorny problem in Einstein's
theory of general relativity. Much attention has been devoted for
this problematic issue. Einstein (E) was the first to construct a
locally conserved energy-momentum complex \cite{1}. After this
attempt, many physicists including  Tolman (T) \cite{13},
Landau-Lifshitz (LL) \cite{3}, Papapetrou (P) \cite{4}, Bergmann
(B) \cite{5} and Weinberg (W) \cite{6} introduced different
definitions for the energy-momentum complex. These definitions
only giving meaningful results if the calculations are preformed
in Cartesian coordinates. Some interesting results obtained
recently lead to the conclusion that these prescriptions give the
same energy distribution  for a given space-time
\cite{V1}-\cite{V8}. Aguirregabiria, Chamorro and Virbhadra
\cite{ACV} showed that the five different energy-momentum
complexes (ELLPBW)  give the same result regarding the energy
distribution with any Kerr-Schild metric. Recently, Virbhadra
\cite{1999} investigated whether or not these definitions (ELLPBW)
lead to the same result for the most general non-static
spherically symmetric metric and found they disagree.
\par
 M{\o}ller \cite{Mc} introduced  a consistent expression for an
energy-momentum complex which could be utilized to any coordinate
system. Some results recently obtained \cite{8}-\cite{11} sustain
that the M{\o}ller energy-momentum complex is a good tool for
obtaining the energy distribution in a given space-time. Lessner
\cite{12} gave his opinion that the M{\o}ller definition is a
powerful concept of energy and momentum in general relativity.
Therefore, it is interesting and important to obtain energy
distribution using M{\o}ller's prescription. Some interesting
results \cite{Xulu1,-3,ACV,Xulu2,-4,I-R,-8,Gad1} led to the
conclusion that in a given space-time, such as: the
Reissner-Nordst\"{o}rm, the de Sitter-Schwarzschild, the charged
regular metric and the stringy charged black hole, the energy
distribution according to the energy-momentum complex of M{\o}ller
is different from of Einstein. But in some specific cases
\cite{1,Xulu1,V97,1999} (the Schwarzschild, the
Janis-Newman-Winicour metric) have the same result. Gad
\cite{Gad3} has shown that the energy and momentum densities of a
G\"{o}del-type space-time in the Landau and Lifshitz and M{\o}ller
prescriptions not give the same result.

\par
In this paper we calculate the energy and momentum densities for
the Weyl metric as well as Curzon metric using M{\o}ller's
energy-momentum complex and compare the result with those already
obtained using energy-momentum complexes of Einstein, Papapetrou,
Landau and Lifshitz, and Bergmann  (see \cite{Gad4}).

\par
Weyl \cite{W1}, \cite{W2} (see also Synge \cite{S}) showed that
static, axially-symmetric gravitational  fields can be expressed
by the metric
\begin{equation} \label{1}
ds^2 = e^{2\lambda} dt^2 - e^{2(\nu -\lambda )}(dr^2 + dz^2) - r^2
e^{-2\lambda} d\phi^2
\end{equation}
where $\lambda$ and $\nu$ are functions of $r$ and $z$ satisfying
the relations
$$
\lambda_{rr} + \lambda_{zz} + r^{-1}\lambda_{r} = 0
$$
and
$$
\nu_{r} = r(\lambda^{2}_{r} - \lambda^{2}_{z}), \qquad \nu_{z} =
2r\lambda_{r}\lambda_{z}.
$$

\par
 For the above metric the determinant of the metric tensor
and the contravariant components of the tensor are given,
respectively, as follows
\begin{equation}\label{3}
%\begin{split}
\begin{array}{ccc}
det (g) & = - r^2e^{4(\nu - \lambda)},\\
g^{00} & = e^{-2\lambda}, \\
g^{11} & = - e^{2(\lambda - \mu)},\\
g^{22} & = -\frac{1}{r^2}e^{2\lambda},\\
g^{33} & = - e^{2(\lambda - \nu)}.
\end{array}
\end{equation}

%\newpage
\section{\bf{Energy-Momentum Complexes}}
The conservation laws of matter plus non-gravitational fields for
physical system in the special theory of relativity are given by
\begin{equation}\label{EM1}
T_{\nu,\mu}^{\mu} \equiv \frac{\partial T_{\nu}^{\mu}}{\partial
x^{\mu}} = 0,
\end{equation}
where $T_{\nu}^{\mu}$ denotes the symmetric energy-momentum tensor
in an inertial frame.
\par
 The generalization of equation (\ref{EM1})
in the theory of general relativity is written as
\begin{equation}\label{EM2}
T_{\nu;\mu}^{\mu} = \frac{1}{\sqrt{-g}}\frac{\partial}{\partial
x^{\mu}}\big(\sqrt{-g}
T_{\nu}^{\mu}\big)-\Gamma_{\nu\lambda}^{\mu}T_{\mu}^{\lambda} = 0,
\end{equation}
where $g$ is the determinant of the metric tensor
$g_{\mu\nu}(x)$.
\par
The conservation equation may also be written as
\begin{equation}\label{EM3}
\frac{\partial}{\partial x^{\mu}}\big(\sqrt{-g}T_{\nu}^{\mu}\big)=
\xi_{\nu},
\end{equation}
where
$$
\xi_{\nu} = \sqrt{-g}\Gamma_{\nu\lambda}^{\mu}T_{\mu}^{\lambda}
$$
is a non-tensorial object and it can be written as
\begin{equation}\label{EM4}
\xi_{\nu} = -\frac{\partial}{\partial
x^{\mu}}\big(\sqrt{-g}t^{\mu}_{\nu}\big).
\end{equation}
where $t^{\mu}_{\nu}$ are certain functions of the metric tensor
and its first derivatives.
\par
 Now combining equation (\ref{EM4}) with equation (\ref{EM3}) we
 get the following equation expressing a local conservation
 law:
 \begin{equation}\label{EM5}
 \Theta^{\mu}_{\nu,\mu} = 0,
 \end{equation}
 where
 \begin{equation}\label{EM6}
 \Theta_{\nu}^{\mu} = \sqrt{-g}\big( T_{\nu}^{\mu} +
 t^{\mu}_{\nu}\big)
 \end{equation}
 which is called energy-momentum complex since it is a combination
 of the energy-momentum tensor $T_{\nu}^{\mu}$, of matter and all non-gravitational field, and a pseudotensor
 $t^{\mu}_{\nu}$ which describes the energy and momentum of the
 gravitational field itself.\\
 Equation (\ref{EM6}) can be written as
 \begin{equation}
 \Theta_{\nu}^{\mu} = \chi^{\mu\lambda}_{\nu,\lambda},
 \end{equation}
 where $\chi_{\nu}^{\mu\lambda}$ are called superpotentials and are
 functions of the metric tensor and its first derivatives.

 %\newpage
\section{\bf{M{\o}ller's Prescription}}

The energy-momentum complex of M{\o}ller in a four dimensional
background is given as \cite{1}
\begin{equation}\label{4.1}
\Im^k_i = \frac{1}{8\pi}\chi ^{kl}_{i,l},
\end{equation}
where the antisymmetric superpotential $\chi^{kl}_i$ is
\begin{equation}\label{4.2}
\chi^{kl}_i = - \chi^{lk}_i = \sqrt{-g}\big( \frac{\partial
g_{in}}{\partial x^m} - \frac{\partial g_{im}}{\partial x^n}\big)
g^{km}g^{nl},
\end{equation}
$\Im^0_0$ is the energy density and $\Im^0_{\alpha}$ are the
momentum density components. \\
Also, the energy-momentum complex $\Im^{k}_{i}$ satisfies the
local conservation laws:
\begin{equation}\label{4.3}
\frac{\partial\Im^k_i}{\partial x^k} = 0
\end{equation}

For the line element (\ref{1}), the only non-vanishing components
of $\chi^{kl}_{i}$ are
\begin{equation}\label{4.4}
%\begin{split}
\begin{array}{ccc}
\chi^{01}_{0} & = 2r\lambda_{r},\\
\chi^{03}_{0} & = 2r\lambda_{z}.
\end{array}
\end{equation}
Using these components  in equation (\ref{4.1}), we get the energy
and momentum densities as following
\begin{equation}
\Im_{0}^{0} = \frac{1}{4\pi}\big(r\lambda_{rr} + r\lambda_{zz}
+\lambda_{r} \big).
\end{equation}
\begin{equation}
\Im_{\alpha}^{0} = 0.
\end{equation}
Using the relation, given below equation (\ref{1}), we found
$$
\Im_{i}^0 = 0.
$$
Therefore, the energy and momentum components are vanishing
everywhere.
 We now restrict ourselves to
the particular solutions of  Curzon metric \cite{Curzon} obtained
by setting
$$
\lambda = - \frac{m}{R} \qquad and \quad \nu = -\frac{m^2
r^2}{2R^4}, \qquad R = \sqrt{r^2 + z^2}
$$
in equation (\ref{1}). \\
For this solution it is found from equation (\ref{4.4}) that the
non-zero components of $\chi^{il}_{j}$ take the form
\begin{equation}\label{}
%\begin{split}
\begin{array}{ccc}
\chi^{01}_{0} & = \frac{2mr^2}{R^3},\\
\chi^{03}_{0} & = \frac{2mrz}{R^3}.
\end{array}
\end{equation}
Using these components the energy and momentum densities for the
Curzon solution become
\begin{equation}\label{3.9}
\Im^{0}_{0} = 0,
\end{equation}
\begin{equation}\label{3.10}
\Im_{\alpha}^ 0 = 0.
\end{equation}
The energy and momentum components are vanishing everywhere.

\par
 In the following table we summarize our results
obtained  (see also \cite{Gad4}) for the energy and momentum
densities associated with  Curzon metric which have been obtained
by using Einstein, Landau and Lifshitz, Papapetrou, and Bergmann
energy-momentum complexes.
\newpage
\begin{table}
  \centering
  \begin{tabular}{|c|c|c|}
  %\begin{tabular}{|t|l|}
    \hline
    {\bf{Prescription}}& {\bf{Energy density}} & {\bf{Momentum density}} \\
    \hline
     Einstein & $\theta^{0}_{0} = \frac{1}{16\pi}\Big[-\frac{4m^2r^2}{R^6} +
\frac{4m^2}{R^4} + $&\\
     &
$2e^{2\nu}\big( -\frac{m^2}{R^4} +
\frac{2m^2r^2}{R^6}\big)\Big]$ & $\theta_{\alpha}^{0} = 0$.\\
     \hline
    Landau and Lifshitz& $L^{00} = \frac{1}{8\pi}e^{4\nu - 4\lambda}
    \Big[-\frac{2m^2}{R^4}+\frac{4m^2r^2}{R^6}-\frac{2m}{R^3}+$ &\\
&$e^{-2\nu}\Big( -\frac{5m^2}{R^4} - \frac{4m^2r^2}{R^6}-
\frac{2m^4r^2}{R^8}+\frac{8m^3r^2}{R^7}+\frac{2m}{R^3}\Big)\Big]$    & $L^{\alpha 0} =0$\\
    \hline
    Papapetru & $\Omega^{00} = \frac{1}{16\pi}\Big[-e^{2\nu - 4\lambda}\big(
\frac{4m^4r^2}{R^8} + \frac{12m^2}{R^4} - \frac{16m^3r^2}{R^7} +$&\\
& $\frac{4m^2}{R^6}\Big) +
2e^{2\nu}\Big(\frac{2m^2r^2}{R^6}-\frac{m^2}{R^4} \Big)\Big]$ &$\Omega^{\alpha 0} = 0$. \\
    \hline
    Bergmann& $B^{00}= \frac{me^{-2\lambda}}{8\pi R^{3}}\Big[ -\frac{2m}{R} +
    \frac{2m^{2}r^2}{R^4} -(e^{2\nu} -1)-$&\\
    & $\frac{2mr^2}{R^3}- \frac{me^{2\nu}}{R} + \frac{2mr^2e^{2\nu}}{R^3}\Big]$&$B^{\alpha 0}=0$\\
    \hline
  \end{tabular}
  \vspace{2mm}
  \caption{\sf{The energy and momentum densities, using (ELLPB),
   for the Curzon metric}}
\end{table}

\section*{\bf{Discussion}}
It has been remained a controversial problem whether energy and
momentum are localizable or not. There are different opinions on
this subject, contradicting the viewpoint of Misner et al.
\cite{MTW} that the energy is localizable only for spherical
systems. Cooperstock and Sarracino \cite{CS} argued that if the
energy localization is meaningful for spherical systems then it is
meaningful for all systems. Bondi \cite{B} expressed that a
non-localizable form of energy is inadmissible in relativity and
its location can in principle be found.
\par
 Using different definitions of energy-momentum complex, several
authors  studied the energy distribution for a given space-time.
Most of them restricted their intention to the static and
non-static spherically symmetric space-times. Using Einstein's
energy-momentum complex, Rosen and Virbhadra \cite{RV} calculated
the energy and momentum densities of non-static cylindrically
symmetric empty space-time. They found that the energy and
momentum density components turn out to be non-vanishing and
reasonable.
\par
 In this paper, we
calculated the energy and momentum density components for the Weyl
metric  using M{\o}ller's prescription. In addition, using these
results we obtained the energy and momentum density components for
the Curzon metric. We compared these results with those obtained
using the four different energy-momentum complexes (ELLPB)
\cite{Gad4}. In the case of Weyl metric it is found that the
energy-momentum complex of M{\o}ller do not provide the same
results for the energy density, comparing with the four different
energy-momentum complexes (ELLPB). On the other hand, we found
that the energy-momentum complex of M{\o}ller for the energy
density associated with Curzon metric  agrees with complexes
(ELLPB) only when $R \rightarrow \infty$. Finally, in the case of
Curzon metric we see that the energy in the  prescriptions (ELLPB)
diverges at the singularity ($R = 0$), but it will never diverge
in M{\o}ller's prescription. 
\end{document}